# A Security Assessment tool for Quantum Threat Analysis

Basel Halak[1], Cristian-Sebastian Csete[1], Edward Joyce[1], Jack Papaioannou[1], Alexandre Pires[1], Jin Soma[1], Betul Gokkaya[1] and Michael Murphy[2]

[1]Electronics and Computer Science School, University of Southampton, UK

[1]ARQIT Security Solution, UK

Email (basel.halak@soton.ac.uk)

***Abstract***. The rapid advancement of quantum computing poses a significant threat to many current security algorithms used for secure communication, digital authentication, and information encryption. A sufficiently powerful quantum computer could potentially exploit vulnerabilities in these algorithms, rendering data in transit insecure. This threat is expected to materialize within the next 20 years. Immediate transition to quantum-resilient cryptographic schemes is crucial, primarily to mitigate store-now-decrypt-later attacks and to ensure the security of products with decade-long operational lives. This transition requires a systematic approach to identifying and upgrading vulnerable cryptographic implementations. This work developed a quantum assessment tool for organizations, providing tailored recommendations for transitioning their security protocols into a post-quantum world. A comprehensive web application, supported by an integrated database and hosting infrastructure, was built to facilitate this solution. It employs a risk calculation formula that evaluates an organization's vulnerabilities based on their responses, providing recommendations derived from extensive research into current systems and encryption capabilities. The work included a systematic evaluation of the proposed solution using qualitative feedback from network administrators and cybersecurity experts. This feedback was used to refine the accuracy and usability of the assessment process. The results demonstrate its effectiveness and usefulness in helping organizations prepare for quantum computing threats. The assessment tool is publicly available at (https://quantum-watch.soton.ac.uk).

1. Introduction

The accelerated development of quantum computing poses a significant risk to prevailing security algorithms, which safeguard data communication, authentication, and encryption. With sufficient computational power, quantum computers could soon compromise encrypted data in transit, exposing current cryptographic vulnerabilities [1-7]. Despite these looming threats, basic cybersecurity practices are frequently ignored until a security breach highlights the oversight, often exacerbated by industry ignorance and apathy.

This negligence often stems from the prohibitive upfront costs of implementing robust security measures, perceived as lacking direct return on investment, particularly as increased security measures diminish perceived threats and financial impacts of potential attacks [8, 9]. Moreover, the trade-off between security and usability can deter the adoption of stringent security protocols [10-11], which may complicate or frustrate user experience. Additionally, a pervasive lack of awareness regarding emerging and existing technological threats often leaves businesses ill-prepared, exposing them to significant data breaches.

The 2023 Cyber Security Breaches Survey in the United Kingdom illustrates these challenges, noting that despite 32% of businesses experiencing breaches or attacks within the past year

[1], cybersecurity remains inadequate across many sectors, particularly within smaller enterprises lacking the resources to adapt to fast-evolving threats [12-37].

Against this backdrop, the proposed research focuses on developing a tool aimed at enhancing the quantum threat awareness among network administrators and technology officers. This initiative seeks to bridge the current knowledge gap by offering an accessible evaluation tool that not only educates users on their systems' vulnerabilities to quantum computing threats but also guides them towards adopting quantum-safe security measures.

This work comprises the development, evaluation, and deployment of this tool, including extensive research into security vulnerabilities posed by quantum computing and the exploration of quantum-safe cryptographic transitions. The research aims to deliver a web application that provides organizations with a robust evaluation of their security posture against quantum threats, culminating in a risk assessment and tailored recommendations. This evaluation relies heavily on qualitative feedback from network administrators and cybersecurity experts to refine the tool's accuracy and usability.

The remainder of the paper is structured as follows. Section 2 provides a summary of vulnerable cryptographic systems and a list of emerging solutions. Section 3 presents the design and implementation methods of the proposed tool. Evaluation approach is discussed in section 4, finally conclusions are drawn in section 5.

## 2. Related Work
### 2.1. Quantum Vulnerable Cryptography
#### a. Symmetric Cryptography

In the domain of symmetric cryptography, encryption algorithms such as the Advanced Encryption Standard (AES), Secure Hash Algorithms (SHA), exhibit vulnerabilities to quantum threats. AES (which was adopted by NIST in 2000) contains three versions of the Rijndael block cipher with varying key lengths: AES-128, AES-192, and AES-256 [38]. These block ciphers have become the foundations for many cryptographic services, particularly ensuring data confidentiality. Due to Grover's algorithm and its reduction in the complexity of the key space of AES it is not considered secure or suitable in a post-quantum world however with a large enough key space and size such as in AES- 256 the algorithm demonstrate much stronger quantum resistance. Aes-256 possible to break with Grover's algorithm, but it would require 3,000 to 7,000 logical qubits in implementation and quantum computers of this scale are likely to be very far away [34, 37].

The Secure Hashing Algorithm (SHA) is a collection of hash functions, the first released in 1995 by NIST as part of the U.S. Government's Capstone project. Hash functions are deterministic and usually irreversible, converting input data to a fixed-size string. This string is used after hashing data and certificates to ensure data integrity, security, and authenticity [39]. The latest member of the SHA family of hashing functions is the SHA-3 release; however, despite the improvements over SHA-2, SHA-256 remains widely used. Nevertheless, both are potentially weakened through the use of quantum computers due to the application of Grover's algorithm[3]. Grover's algorithm allows quantum computers to be much more efficient than classic computers at finding hash collisions with a quadratic speed improvement. Quantum computers second advantage is in a preimage attack brute force attack where they can cover the key space of a set key size (n) in $2^{n/2}$ compared to conventional computers $2^n$ [40]. SHA-512 currently exhibits the highest resistance to quantum threats due to its very large solution space size. Figure 1 compares the time complexity for a collision attack on hash function different hash sizes (see figure 2 form details).

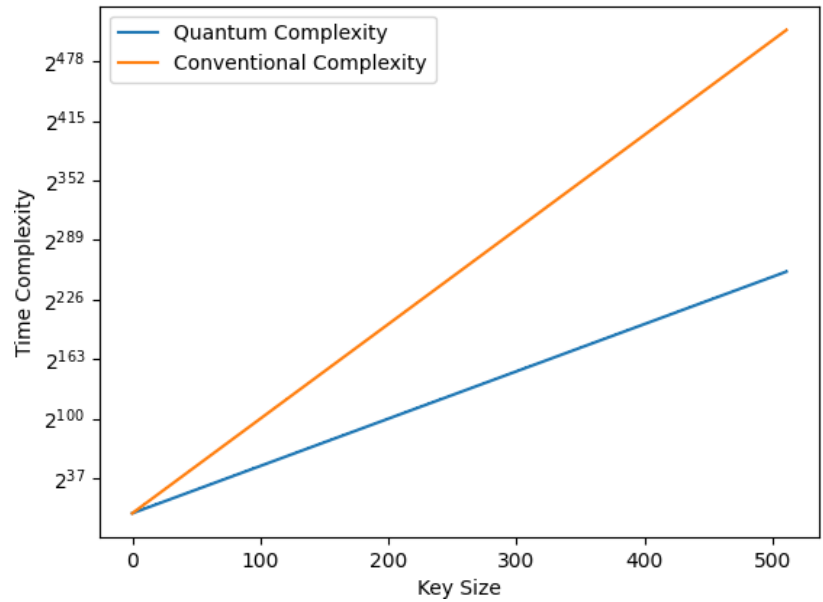

Figure 1: Comparing Quantum and Conventional Time Complexity for Hash Sizes

| Hash Function | Depth | Width |
|---|---|---|
| MD5 | 29,252,464 | 801 |
| SHA-1 | 36,475,332 | 2913 |
| SHA-224 | 38,953,918 | 2593 |
| SHA-256 | 38,953,918 | 2593 |
| SHA-384 | 99,089,086 | 6209 |
| SHA-512 | 99,089,086 | 6209 |
| SHA-512/224 | 99,089,086 | 6209 |
| SHA-512/256 | 99,089,086 | 6209 |
| SHA3-224 | 1,131,675 | 2161 |
| SHA3-256 | 1,131,675 | 2193 |
| SHA3-384 | 1,131,675 | 2193 |
| SHA3-512 | 1,505,375 | 2449 |
| SHAKE128-256 | 1,131,675 | 2193 |
| SHAKE256-512 | 1,505,375 | 2449 |

Figure 2: Cost of Preimage for Various Hash Functions (Reproduced from [3])

b. *Asymmetric Cryptography*

Several asymmetric cryptography algorithms, including the Rivest-Shamir-Adleman (RSA), Elliptic Curve Cryptography (ECC), as well as the Diffie–Hellman key-exchange pro- tocol, are less effective when targeted by quantum computing threats.

RSA is commonly used to secure browsers and web server connections to ensure privacy and integrity of data-in-transit over the Internet [28]. Shor's algorithm can effectively break RSA-2048 encryption if given sufficient qubits [4, 37] as RSA is based on the prime factorisation problem, which is made more trivial by Shor's. Figure 3 reveals 87.5% of experts believe that RSA-2048 will be decrypted in less than 24 hours within the next 30 years due to the rate of growth of qubits per year.

ECC is a cryptographic protocol based on the discrete logarithm problem, introduced in the mid-1980s by Koblitz and Miller [41]. Instead of utilising the product of large prime numbers to generate keys, ECC uses elliptic curve equations for key generation. A significant advantage of this approach lies in its efficiency in achieving comparable security with substantially shorter key lengths. For instance, the security offered by a 1024-bit RSA key is equivalent to a 163-bit ECC key, an 84% reduction in key size [41]. However, ECC shares a vulnerability to quantum computing threats with RSA, its security foundation lies in discrete logarithms, rendering it susceptible to Shor's algorithm as well [27, 23].

The Diffie-Hellman protocol, developed by Whitfield Diffie and Martin Hellman in 1976, was among the earliest methods proposed for key exchange. It formed the foundation for subsequent cryptography, such as ECC, by introducing new cryptographic systems capable of securely exchanging keys over a public channel. It reduces the dependence on secure key distribution channel methods with a pre-existing shared secret [42]. Diffie- Hellman is also vulnerable to emerging quantum threats as it is rooted in the discrete logarithm problem, and Shor's algorithm is therefore effective against it. Consequently, the protocol is not suitable for a post-quantum future, and alternative key exchange methods must be used.

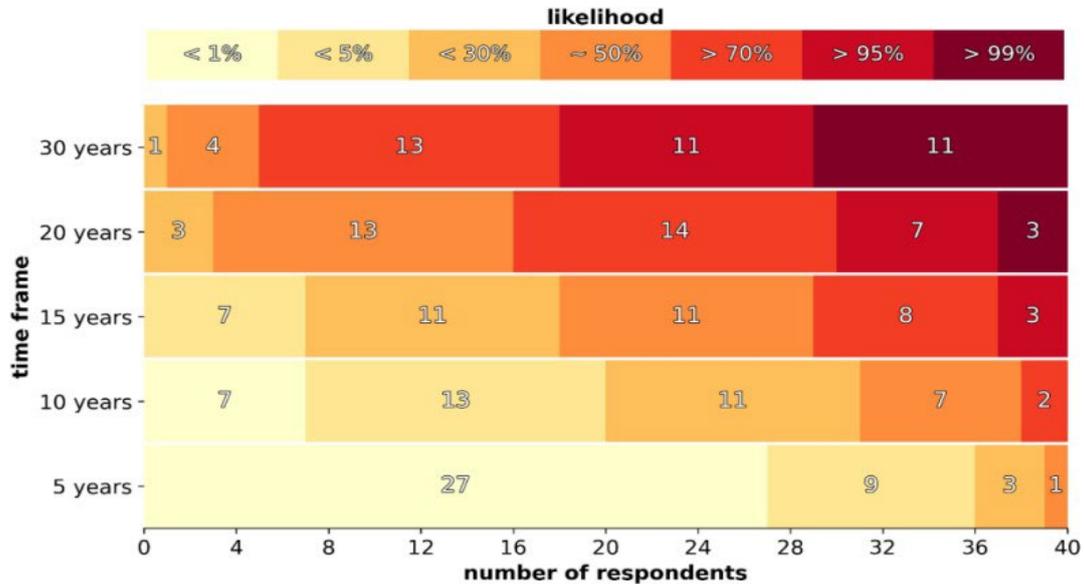

Figure 3: Likelihood to Break RSA-2048 in 24 Hours (Reproduced from Global Risk Institute [4]).

*c. Security Protocols*

This work has identified a set of protocols that will be affected by a quantum computer attack, shown in Table 1. This is not an exhaustive list, and can be expanded with other existing or emerging schemes that employs quantum-attack susceptible cryptographic algorithms. TLS, which contains symmetric and asymmetric cryptographic techniques, is a security protocol widely employed in digital communications. TLS encrypts connections through a mechanism known as the "handshake" event. This process facilitates message exchanges between the client and the hosting server by establishing a secure connection using asymmetric encryption to generate a shared key. This key is then used as part of symmetric encryption for the remainder of communication, as it is computationally lighter and faster than asymmetric encryption [43]. Given that TLS is usually configured with RSA, AES-128, Diffie Hellman and elliptic curve cryptography, the current protocol cannot be considered secure against quantum threats and extensions need to be made to improve the security measures in newer versions, such as Cloudflare's KEMTLS project [44]. Future adversaries possess the potential to undermine the confidentiality of Diffie-Hellman by utilising Shor's algorithm to decrypt the key exchanged during TLS initiation [5]. Furthermore, authenticity can be jeopardised due to the vulnerability of asymmetric encryption, which is essential to the security of certificate signatures in Public Key Infrastructure (PKI).

Table 1: Security Protocols Vulnerable to Shor's Algorithm

| Protocol | Description |
|---|---|
| SSL | Secure Sockets Layer |
| TLS | Transport Layer Security |
| SSH | Secure Shell |
| DNSSEC | Domain Name System Security Extensions |
| WPA2 | Wi-Fi Protected Access 2 |

## 2.2. Emerging Quantum Safe Cryptographic Solutions

### a. Post Quantum Cryptography

The national institute for security and standards (NIST) is currently collecting submissions and promoting research to establish standard- ised public-key cryptographic algorithms to mitigate the quantum threats of the future [6]. Chosen algorithms are held to high-security standards, requiring forward secrecy, resistance to side-channel and multi-key attacks, and unforgeable digital certificates [6]. Performance and cost metrics are taken into account, such as the computational efficiency of key generation, the transmission expenses for public keys, signatures, or cipher texts, and the hardware resource requirements for algorithm implementation [6]. Flexible, simple, and adaptable algorithms were prioritised for smooth implementation and widespread adoption.

CRYSTALS-Kyber was chosen as the key encapsulation mechanism (KEM), and for digital signatures, the selected algorithms include CRYSTALS-Dilithium, FALCON, and SPHINCS+. NIST recommends CRYSTALS-Kyber for key establishment and CRYSTALS-Dilithium for digital signatures [6]. CRYSTALS-Kyber is lattice-based cryptography and distinguishes itself in rapid key generation, encapsulation, and de- capsulation. Kyber's efficiency and strong performance metrics in both software and hardware implementations make it an optimal choice for standardisation.

Lattice-based cryptography is an area of developing research in cryptography that provides hard-to-solve mathematical problems for both classical and quantum computers. Not- able examples include CRYSTALS-Kyber and CRYSTALS-Dilithium. The core of this cryptography lies in the complexity of solving lattice problems, mathematical structures formed as an infinite collection of points regularly spaced out in a grid-like formation. Learning with Errors (LWE) takes advantage of this complexity with the task of differentiating between noisy linear equations and random vectors within these lattices, making deciphering computationally infeasible even with a quantum computer [45, 46]. In practical applications, lattice-based cryptography is utilised in securing digital signatures, as seen in systems such as CRYSTALS-Dilithium, FALCON, and SPHINCS+ [6].

### b. Arqit QuantumCloud™:

This is a commercial solution developed by Arqit, QuantumCloud™ to provide a quantum-safe solution to enhance secure communication via a Platform-as-a-Service software. It employs a zero-trust model by re-authenticating each endpoint during interactions and validating permissions continuously. The protocol aligns with NIST recommendations, maintaining forward secrecy by dynamically transforming keys with each authentication. Key generation is client-side, using materials from QuantumCloud's service, ensuring the security of the final keys, which are AES-256 based and meet the size recommendations for quantum resistance. Recent studies highlight its efficiency, noting a significant reduction in energy consumption compared to traditional systems.

### c. IBM Quantum Safe Technology (QST)

IBM offers a comprehensive package aimed at facilitating a quantum-safe transition through its Quantum Safe Technology suite. This suite includes tools like Quantum Safe Explorer, which scans for cryptographic vulnerabilities; Advisor, which builds a cryptographic inventory; and Remediator, which assists in redesigning system architectures to support post-quantum cryptography (PQC). Despite its robust capabilities, the accessibility and cost of IBM's QST can be prohibitive, limiting its practical use to a broader audience.

This research underscores the necessity of lowering barriers to entry for quantum-safe transitions by introducing accessible, cost-effective tools that raise awareness and facilitate the adoption of secure technologies against quantum threats. The development of such tools, including the proposed publicly accessible web application, is critical in preparing organizations for the future cybersecurity landscape, ensuring they are equipped to handle emerging quantum computing risks effectively.

## 3. Tool development Method
### 3.1. Questionnaire design

The formulation of the assessment tool's questions was conducted through a multi-stage process, incorporating iterative feedback from customers, security experts, and internal testing. This iterative methodology was crucial for enhancing the scope, applicability, and clarity of the questions. As the project progressed, the methodologies employed in question development evolved in response to extensive research. The domains of the questions were segmented into distinct sections utilizing three diverse strategies, each providing unique advantages. This process necessitated meticulous compilation and review at each phase to ensure the integration of constructive feedback and the refinement of the question set. Questions were maintained in two primary repositories: the development database, which stored questions used during application testing, and the Question Master Spreadsheet (QMS). The QMS served as a collaborative platform where questions could be readily reviewed and revised prior to their incorporation into the database. This spreadsheet included all essential information for each question, such as risk scores and recommendations.

The OSI model, a recognized standard in computing that categorizes network communications into seven layers, provided a foundational framework for question formulation. This model was particularly relevant as it aligns directly with the primary target of quantum threats—data in transit—which can be intercepted and decrypted. Utilizing the OSI model facilitated a structured breakdown of the network stack, allowing for focused research and categorization of potential vulnerabilities and corresponding recommendations. This approach not only captured areas previously overlooked in initial question development phases but also ensured that each question was directly linked to relevant quantum security recommendations, vulnerabilities, or policies. In this context, this work developed table of vulnerabilities for the most widely used applications. Examples of these are included in Table 2 below.

Table 2: Examples of applications vulnerabilities to quantum-computer Attacks

| Application | Encryption method | Quantum vulnerability | Consequences |
|---|---|---|---|
| **Google Chrome** | TLS (using RSA, ECC, DH for key exchange or signatures) *only for version 115 or older | Vulnerable to Shor's algorithm which can factorize RSA and compute discrete logarithms for ECC and DH | Exposure of sensitive data like passwords, credit card details. Risk of man-in-the-middle attacks |
| **Microsoft Teams** | SRTP for voice/video and TLS for instant messages/signaling | Asymmetric components (e.g., RSA or ECC) of key establishment can be compromised by quantum algorithms | Unauthorized interception of business meetings and communications. Confidentiality breaches leading to business damages |
| **OneDrive** | TLS for data in transit (using RSA, ECC, DH) | Vulnerable to Shor's algorithm in its asymmetric components | Unauthorized access to files, potential legal repercussions from data leaks, and loss of intellectual property |
| **Google Cloud** | Uses TLS for data in transit (IPSec, S/MIME for mail | Can be cracked by Shor's algorithm 1994 | Unauthorized access to confidential business information, business files, application source code, private emails, business account breaches, personal data access etc. |
| **Slack** | Includes the use of TLS 1.2 for data in transit | Can be cracked by Shor's algorithm 1994 | Can lead to unauthorized access to confidential information within messages shared by employees aswell as business accounts |
| **Zoom** | .2, AES 256-bit GCM | | |
| **AWS** | TLS 1.2/1.3 | https://aws.amazon.com/blogs/security/post-quantum-tls-now-supported-in-aws-kms/ TLS supported? | er information, business operations, and proprietary applications - could be accessed or tampered with. Can lead to data breaches, service interruptions, or malicious changes to business applications. |
| **Salesforce** | TLS 1.3 | | customer profiles, sales data, and business forecasts compromised. Can undermine the trust between a company and its clients, and expose the company to regulatory fines for violating data protection laws. |
| **Dropbox** | TLS 1.2 | | n theft, financial losses, or unauthorized publication of sensitive materials. |
| **rello (low priority)** | TLS 1.2 | | ompetitors insights into a company's upcoming products, strategies, or potential vulnerabilities. |

In addition to the OSI model, network diagrams were employed as a technique to visually represent and understand network configurations, facilitating the intuitive comprehension of complex network topographies. These diagrams typically illustrate the devices, connections, and services within an organization's network, providing valuable context for developing questions that are closely aligned with actual business network setups and their security challenges.

This method complemented the OSI model-based approach by covering aspects of the network that are less about data encryption and more about overall network architecture and vulnerabilities, thus enriching the depth and breadth of the security assessment. Network diagrams also served as a practical tool in the internal testing phase, aiding in the simulation of network configurations to verify the accuracy and relevance of the formulated questions.

The dual use of the OSI model and network diagrams ensured a comprehensive approach to the development of assessment questions, catering to a detailed understanding of both general network functions and specific security vulnerabilities pertinent to quantum threats. This robust methodology underpins the tool's effectiveness in equipping organizations with the necessary insights to fortify their defenses against emerging quantum computing challenges.

Additionally, a help pop-up was provided for almost every question. This information guides users to where they can find information either on their devices or online as well as explaining some of the technologies and network elements that a user may not be familiar with such as a Cryptographic Center of Excellence which has the following help pop-up.

A full list of questions is included in Appendix 1

### 3.2. Result Formulation Methodology

The completion of the questionnaire yielded two main outputs: risk categorization and recommendations. The following two sections will discuss the rationale behind their presentation.

#### a. Risk Scores

Risk score is a representation of the diagnosed system's potential exposure to cybersecurity threats based on the provided answers. It was defined as a value ranging from 0 to 3, where 0 indicates almost no risk and 3 represents high risk exposure. Each answer choice was manually assigned a risk score and summed for the output of the prototype to signify overall system risk. For multiple-choice typed questions, the scores of selected answers were added to represent its risk score. The total was divided by the the sum of maximum risk score per question as shown in Equation 1 and 2.

$$Risk = \frac{\sum_{i=1}^{N} risk[i]_{selected}}{total} \quad [1]$$

Where

$$total = \sum_{i=1}^{N} \max(risk\ scores[i]_{[single-choice]}) + \sum_{i=1}^{N} Risk\ Scores[i]_{multiple-choice} \quad [2]$$

The final calculated risk score was categorised into low (0-33), medium (34-59), or high risk (60-). A weakness in this approach lies in risk skewing towards multiple choice questions as it takes its sum. Initial implementation attempted a more sophisticated approach, where risk was defined as the weighted product of probability and impact. This was due to risk being characterised as a combination of breach frequency, likeli- hood, and cost in literature [51, 52, 53]. However, assigning a value to each variable became overly complex when these terms apply at different scales per industry and insufficient information was available to evaluate a numerical value. For instance, 'im- pact' can be broken down into cost inside and outside the organisation. Assessing the direct financial impacts of an attack was difficult, but its social consequences, such as business loss in brand image, proved more challenging to estimate given the limited information the prototype had access to. Calculating every variable led to unreliable broad estimations, needlessly over complicating the process. This meant that while the initial approach seemed more logically sound, it could lead to subjective and ambiguous results. Therefore, for the purpose of this work, highlight quantum computing risks, a simple value-assign approach was deemed more appropriate.

### b. Recommendations

The recommendations were the main customer facing element of this work, they provide the value to the users and provide information on how to transition to post-quantum security as well as educating the user on what the quantum threat entails and what dangers quantum computers pose to their data. Some questions do not contain a re- commendation, sometimes this is because the question is a "chain question" and leads the user to different paths depending on their answer where their initial answer doesn't provide any security information that is relevant.

The recommendations are directly seen and read by users which contrasts with our risk scores which are hidden and only a "risk level" given to the user, because of this the recommendations needed to be accessible and explain security concepts that might not be familiar to our target audience. The recommendations provide information on technologies and techniques that are either available or in active development as well as suggestions for policy or infrastructure alterations that could improve company resilience.

The prototype presents the top five main highest risk recommendations initially for two reasons. Firstly, an excess of suggestions could overwhelm and confuse users due to a lack of clear priority, leading to inaction and undermining the prototype's utility. Limiting recommendations to a manageable number ensures clearer resource allocation and a higher likelihood of implementation. Out of 56 total recommendations, many were considerably minor, such as updating a browser. As such, the Pareto principle, which suggests that a smaller proportion of efforts (~20%) yields the majority of results (~80%), was leveraged to provide conciseness and effectiveness to the end user [54]. In this context, the selected five recommendations —constituting roughly 9% of the total 56 recommendations— represent the most effective measures, striking a delicate balance between clarity and impact.

## 3.3. Application Design

For the development of the web application React and Django were the chosen frame- works. React, which is a JavaScript library, was chosen for the front end development, to facilitate the process of building an interactive user interface and enabling a dynamic and responsive user experience. On the back end, Django, which is a high-level Python web framework, was selected to facilitate rapid development and due to its robustness, scalability, and security. Besides this, Bootstrap was also used to facilitate the development and styling process of the user interface.

The interface was designed with a responsive layout, in order to cater to users accessing the system across various devices, including both desktop and mobile platforms. This is done by adapting the styling and layout to accommodate different screen sizes and types of devices, and, consequently, enhancing accessibility and usability for all users.

Regarding the structure of the web application, it was divided into 4 main different stages. These stages are the introduction, section selection, question answering, and results display. The final three stages of the web application, collectively form an analysis tool that aims to help companies in their transition into post quantum cryptography. All these 4 stages are described in more detail in the following subsections.

### a. Landing Page

The first one, being a start page, as Figure 4 shows, which purpose was to provide as much information as possible on what is the quantum threat, what risks does it pose to the current cryptography infrastructure, and why should companies start to prepare for this emerging threat. To accomplish this, this starting page, which is the page the user sees first when they access the web application, has some statistics in the top, in conjunction with a graph 5 that indicates the increase in the quantum threat level over the next years, as new and more powerful quantum computers are revealed, as well as a video below, to further explain and illustrate what exactly is the quantum threat and how does it endanger the current cryptography methods. On the first page, there is a button to find out more, that scrolls down to the video, and a button to start the analysis process. This button takes the users to the second stage, the section selection. Besides this, there is also a button to resume a previously started analysis that takes the user directly to the third stage, if the user is accessing the web application from the same device and browser as before.

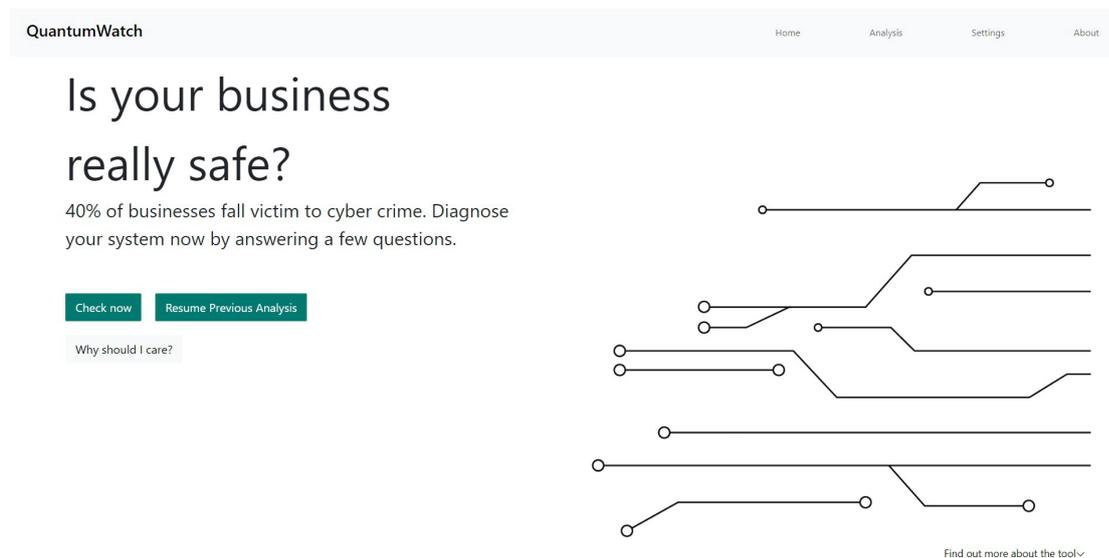

Figure 4: Landing Page

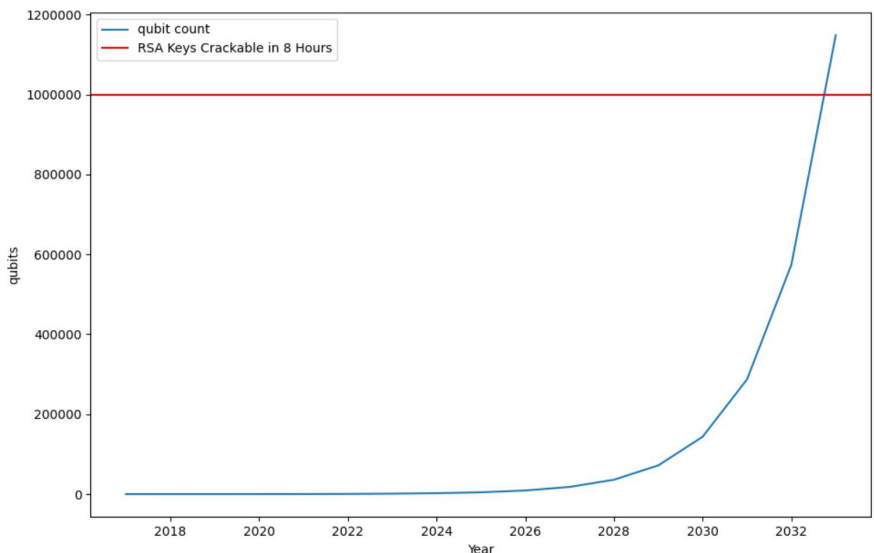
Figure 5: Prediction of Quantum Threat Level

### b. Section Selection

At this stage, the user is presented with a few different sections and should select those that apply to their company or organisation, as shown in Figure 6 To accomplish this, a get request is used to get all the section names from the database. From these, only the non-mandatory sections will be displayed as options that the user can choose from. From the current sections, only the Introduction is mandatory, so, therefore, this is the only section that does not appear as an option at this stage. While the front- end is waiting for the list of section names from the get request, a Bootstrap Border Spinner is displayed, even though the responses are usually fast. After receiving this list of names, the user is prompted with a question on the top of the page, that asks the user to select which options best describe the technology that their organisation relies upon. The options are displayed as squares, with a title on the top left of each square, to represent each section, an image, in the centre, related with that section, and a small description of what each section includes or consists of, in the bottom of the square.

A section selection square component was created, in order to be able to dynamically create these squares. In this stage, the user can select all of the sections that include technology used or relied upon in their organisation, from a minimum of 1 section to a maximum of all the sections available. The user cannot progress to the next stage, until at least one section is selected. If the user tries to progress to next stage by clicking the next button without selecting any sections, then an alert will appear, to remind the user to select at least one relevant section, and the user will not be progressed to the next stage. A React State is used to keep track of the sections the user has selected, and to update them, as the user selects or unselects sections.

On the bottom right, there is a button that saves the selected sections into the local storage of the user's browser, and takes the user into the following stage, once they have selected all the sections that are relevant to their company. For user specific data storage, which includes the users' answers, local storage was chosen to store this data, in order to not save user sensitive data in the database, allow for users to leave, close their tab or computer and be able to return and resume their analysis, given they use the same device and browser as before, and due to the larger data storage limit, when compared to alternatives like cookies.

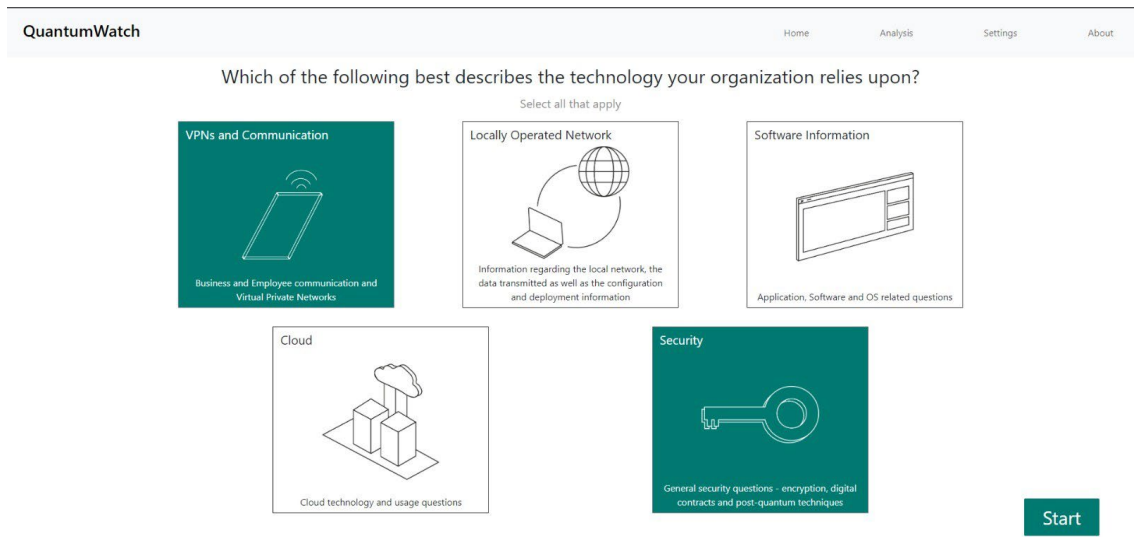

Figure 6: Section Selection Page

### c. Questionnaire

At this stage, the user can see the sections that they need to complete, and their overall progress in the analysis process. An example of this section progress page is shown in Figure 7 Like in the previous stage, while these sections have not been received from the back-end, a Bootstrap border spinner is displayed to indicate the page is not fully loaded yet. At the top, the user can see how many sections they have completed already out of the ones they have to complete, in conjunction with a progress bar that indicates the same but as a percentage. Below that, the user can see a square for each section they selected previously, as well as for sections that are mandatory, like the Introduction section. These represent the sections that the user needs to complete in order to progress to the next stage, the results.

Each of these squares provide the user with the same information for each section as before in the section selection stage, so that includes the title of the section in the top left of the square, a related image in the centre and a small description of what that specific section includes, below the image. Besides this, unlike the section selection, at this stage, each square also mentions the expected time duration that each section should take to complete. This time estimation is in the bottom right corner of each square, and is presented in minutes.

Furthermore, each square also indicates whether that specific section has already been completed by the user or not. When the page is first loaded, the web application starts by checking if any of the sections are marked as complete in the user's browser local storage, in order to know whether to render them as completed or not.

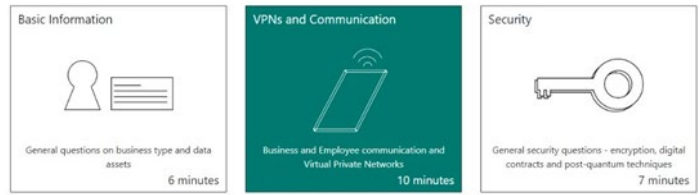

Figure 7: Progress Page

When the user clicks on a section square, they get redirected to the section page, and the web application makes a get request with the id of the section that the user selected, in order to receive the list of questions and possible answers that are part of that specific section, from the database.

All of the questions are either single choice or multiple choice questions. The information received from the database, consists of a list of question objects. Each of these question objects provide the question text, a list of the possible answers for that specific question, question's type, so either multiple choice or single choice, and also if this is a chained question (explained later on), if it is it also mentions the list of answer ids that trigger this question to be shown.

While the web application waits for the list of questions and possible answers from the database, a Bootstrap Border Spinner is displayed at the centre of the page, like in the previous sections.

After the list is received, the section page populates with the entirety of the questions that are part of that section. At the start, the first question is displayed, and the user can immediately navigate through all of the questions by using the navigation buttons displayed on the screen. Figure 8 shows an example of a question in this section page.

The section page has the section name at the top in the centre, the question type just under that, so either single choice or multiple choice, and the question text below that, also in the centre, but with a larger size. Below these, all of the possible answers to this question are displayed as rectangles, with the answer text inside.

If the current question is multiple choice, the user will be able to select as many answers as possible from the displayed options. However, if the current question is single choice, if the user already has an answer selected, and proceeds to select another one, the previous answer will be unselected, and the most recent selection will be displayed as selected.

When the answer rectangles are loaded, the web application starts by checking if any of the answers from this specific question have been previously selected by the user, in order to know if they should be rendered as selected or not. This is accomplished by checking the user's browser local storage to see if the answers are part of the list of selected answers stored there.

Below those possible answers, there is a help button. This button's purpose is to provide further information or even a list of detailed instructions on how to find some technical information necessary, to help the user answer the question. This button, when clicked, triggers a pop up to appear in the centre of the screen with the relevant help information related with the question the user is trying to answer. To accomplish this, a Bootstrap Modal was used to facilitate the pop up creation.

Just under this, there is a progress bar which indicates the number of the current question being displayed out of the whole set of questions for that section. The total number of questions is derived from the size of the list received as an answer at the start, and a React State keeps track of the number of the current question being displayed. This number functions as an index of the current question, in the full questions list.

Below this progress bar, there are three buttons. A previous button that takes the user to the previous question, located in the bottom left of the page, an exit button that exits the section page and goes back to the section progress page, without submitting the user answers, located in the bottom centre of the page.

And finally, a next button that takes the user to the next question, located in the bottom right of the page. This button turns to a submit button, if the current question displayed is the last one of that section.

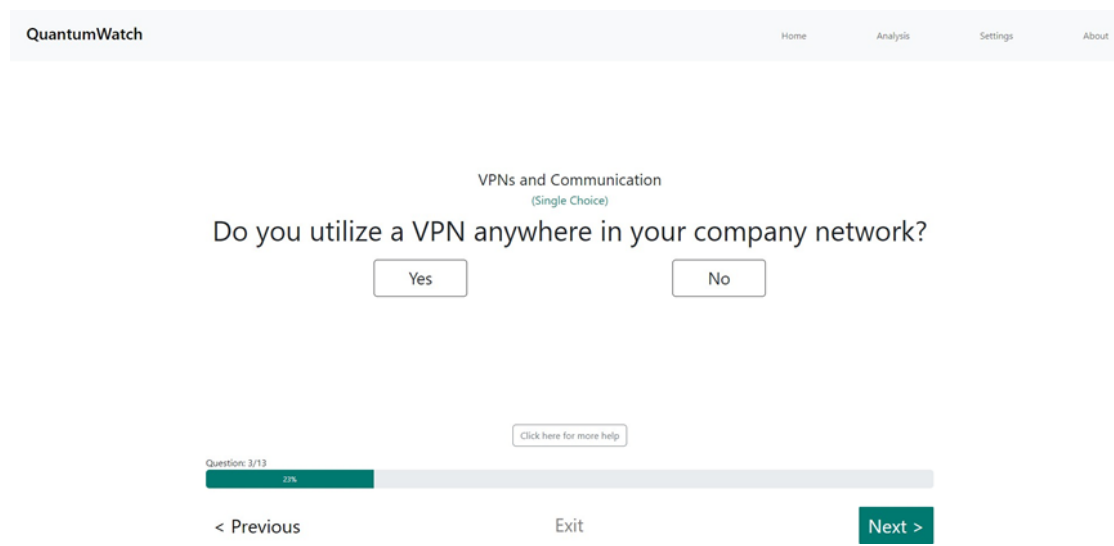

Figure 8: Question Page

The previous and next buttons update the current question number React State men- tioned above, by decreasing or increasing the number by one, accordingly, and then update the page, in order to show the new question and answers. There are some types of questions that are only displayed depending on whether the user selected certain answers to previous questions in that section. These questions are chained questions, in the sense that they are chained to certain answers to previous questions, and only appear if those were selected. The main purpose of these types of questions is avoiding asking non relevant questions. For example, if a company does not use any IoT devices, then they should not have to answer questions regarding their IoT devices characteristics.

To accomplish this, after the user clicks next, if the next question is a chained question, the web application will start by checking whether any of the trigger answers for that chained question was selected previously by the user. The same check applies when the user clicks the previous button, but instead of skipping to the next question, the web application skips to the previous question.

This is done by checking if the React State that handles the answers user has selected, contains any answer id from that question's list of trigger ids, received at the start together with all of the questions and answers. If it contains at least 1, then the question will be displayed, otherwise, the question will be skipped.

Another React State is used to keep track of which button has the user clicked last, between the next and previous buttons. This is used to figure out whether to move to the next question or previous, when a chained question needs to be skipped.

If the user clicked next, more recently than previous, then the web application will skip to the question after the skipped chained question. If the user clicked previous more recently, then the web application will skip into the question before the skipped chained question. Furthermore, it follows the same logic even when there are multiple chained questions one after the other, that need to be skipped.

Furthermore, like before, a React State is used to keep track of the answer the user has selected. And when the user reaches the final question and presses submit, the selected answers will be saved into their browser's local storage, for the same reasons as stated previously, and the section will also be marked as complete in that local storage. This allows the user to access this section again in the future, and change their answers, without having to re-answer all of the section's questions again.

After that, the user gets redirected back to the section progress page, and the section that they just submitted will now be marked as complete. Besides that, the progress bar on top, will also update, to show the current progress. Nonetheless, the user will still be able to access that section again at any time and change their selected answers.

From here, the user can select another section that has not yet been completed, or, if all other sections have already been submitted and the user does not wish to change any more of their answers, then they can go ahead and click the next button at the bottom right of the page, in order to progress into the final stage, which is the results.

However, this button will only be available after the user has submitted all of the sections at least one time. Nonetheless, this does not mean the user needs to answer all questions, since a section can still be submitted if some questions have not been answered yet.

As expected, as more sections are completed by the user, more squares will be marked as completed, the progress bar also updates accordingly, and more user answers will be saved to the local storage. When the user has submitted all of the sections, all of the squares will be marked as complete, and the user will then be given the option to progress to the results page.

### d. Results Dashboard

The final stage of the analysis, as expected, is the results page, as shown in Figure

9 Like in the previous stages, while waiting for the response from the back-end, the page will display a Bootstrap border spinner, that indicates the results are not ready for display yet.

At the top of the results, there is an indication of what risk category does the user answers fit into. As mentioned before, this can be either low, medium, or high, depending on the final risk score calculated based on all the answers the user selected. The exact boundary values, that define each category are discussed in more detail later in the survey improvements section.

Below this, there are three boxes with relevant information. Firstly, the box on the top left, mentions how many recommendations the analysis produced for the user, based on their answers. Also located inside this box, a see more button will take the user to the list of recommendations located below in the page.

Secondly, there is a box on the top right that indicates the risk level of the user's organisation based on their answers. This provides the risk category, and a risk level bar that further illustrates the risk level. Thirdly, the box just below those, has a small explanation of the user's risk category.

Below all of these, at the bottom of the page, the user is provided with a ordered list of relevant recommendations, ranked by importance level. This importance value can range from 0 to 3, so the recommendations with higher importance, in this case with value 3, will be ranked higher in the list. In case the user received more than 5 recommendations, only the top 5 will be displayed. Nonetheless, the user has the option to see a full detailed list of all the recommendations, by clicking the see more button at the bottom of the recommendations list. After clicking this button, a see less button will appear at the end of the full detailed list of recommendations so that the user can return to the initial top 5 recommendations only display.

To accomplish this, when the user first accesses the page, a post request is submitted to the back-end. A list of the answer ids the user selected is provided in the request, and as a response, the final risk value and a list of the relevant recommendations is received from the back-end. This final risk value received is the value used to define the risk category, and the list of recommendations received is the list of recommendations displayed at the results page. Before being displayed, the web application only needs to order them based on their importance level, and then they are ready for display.

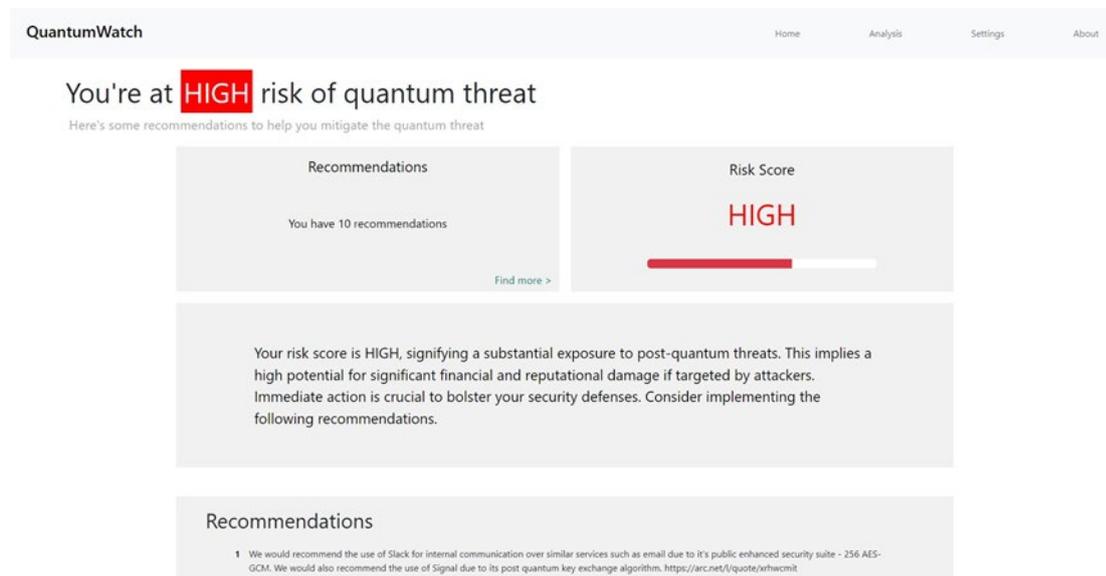

Figure 9: Results Page

### 3.4. Software Development

#### a. Question Database

The first task in deciding the database was to choose between SQL and NoSQL type of database. All members of the team had experience with both type of databases, in the end SQL was chosen due to the fact that the data had well-defined structure and relationships between the different tables. On top of that SQL databases are suitable for read-intensive workloads, due to the type of data stored in the database this brings a beneficial performance advantage.

For the database the default option provided by the Django framework was a local SQLite3 database. This database would not fit the needs of the project because it could only be used locally and not everyone would have the same database during the development process.

A cloud solution was considered afterwards, so that the database is consistent with the whole development team. The decision was to go with Microsoft Azure SQL database. This was a database service that provided a free tier of 100,000 vCore seconds and 32 GB of storage, and the setup involved making a SQL Server, a SQL database and linking it through a connection string with the back end of the application. In the early stages this was a good solution as it provided fast setup for the people working on the development of the application, but as the application grew in the number of requests to the database, the free tier was not enough anymore, even though the speed of the processors were good enough, the amount of transactions was not. On top of that having all the packages up to date to the latest versions, due to security reasons, was not possible because running the application required a driver which was not updated to the latest version on the Mac operating system suite, thus slowing down the development process.

The last two database deployments both used PostgreSQL. This was chosen because it does not require any special drivers to run on the developers' machines and it was also in the beginning the only option when the deployment was on a web hosting service provider.

In Figure 10 there is a representation of the database models and the connections between them. In the beginning there were more models which would store the data that the user would input but due to ERGO regarding what data can be stored, it was decided to not store any kind of sensitive information which might lead to identifying a business. There are also limits on the sizes of the elements in the table, for example, for the question text a limit of 1000 characters was set in order to not overload the database, keeping the good performance, and to allow the database engine to optimize the storage.

Django provided a way to design the application's data through a "code first" approach. This means that the tables in the database were created through the ORM (Object- Relational Mapping) system, which allowed writing code in D.30 of how the tables should look like and then using "makemigration" command would automatically generate a code similar to D.2 which then through the "migrate" command would update the database.

To see what data there is in the database multiple ways were used. One of the meth- ods was to use SQL Server Management Studio and Azure portal interface, when the development had a Microsoft SQL database, and then moving to pgAdmin 4 when the development had a PostgreSQL database. For adding data to the database the Django admin panel was used.

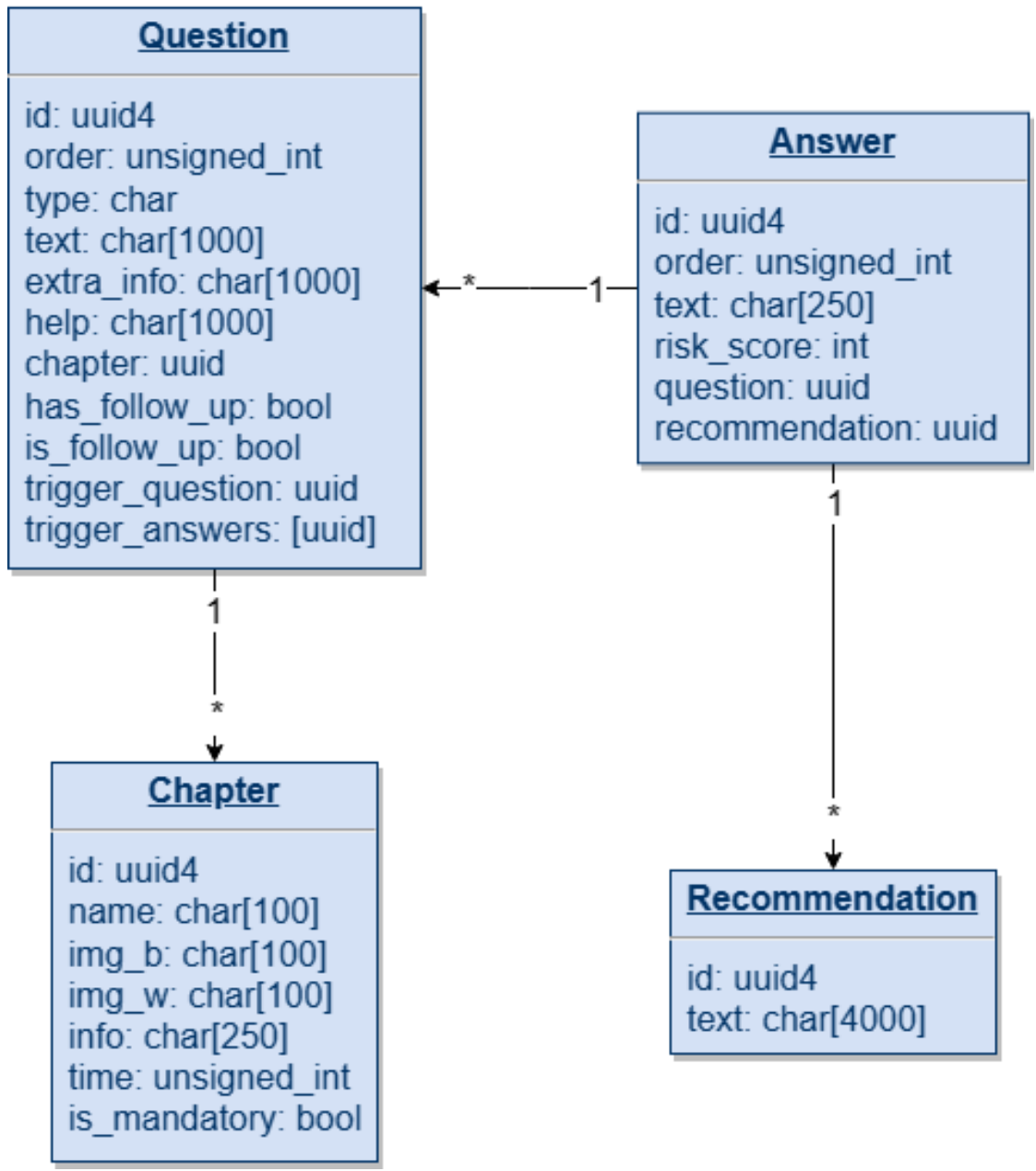

Figure 10: Database Diagram

b. *Data Flow Architecture*

The back end part was built by keeping a REST (Representational State Transfer) architecture in mind. Through making use of API calls the client side can request the chapters, questions for a specific chapter and calculation of the results. The CRUD (Create, Read, Update, Delete) operations on the database were already implemented in the Django admin package.

c.  Deployment

The deployment can be split in two phases, based on where the application was deployed. In the first phase the application was deployed on a web hosting server, Render, which provided free tiers for at least three months for both the database and the web application. The deployment process involved moving the git repository from GitLab to GitHub because there was no available feature to link the GitLab repository on the website. The application also followed some modifications:

- Some packages were downgraded because of the host server capabilities.
- A WSGI HTTP Server (gunicorn) package was installed, and some settings were configured
- The static files were compiled using the command "collect static" provided by Django framework
- Some changes were made to the settings file code  so that the files are served properly and the server redirects the requests properly through gunicorn

The second deployment was on a university virtual machine (VM). The VM setup was handled by the university and the specs for it are as follows:

- System: Red Hat Enterprise Linux release 9.3

- CPU Model: Intel(R) Xeon(R) Silver 4214R CPU @ 2.40GHz

- CPU Cores: 2

- CPU Architecture: x86 64

- RAM: 3.53 GiB

In order to make the web server run on the VM there was a need for a domain name and TLS keys which were provided by the university. These were linked with the python code through NGINX, an HTTP and reverse proxy server, so that all the requests that come to that domain name are redirected internally to the python app which is running continuously on the VM.

4. Evaluation
4.1. Survey Design and Analysis
In this section, the methodology through which user feedback was collected for the web application is explored. The main goal was to obtain better insights into the user's experience, perceptions, and possible recommendations to enhance both the application's usability and the quality of the questions presented along with their relevance to the topics.

To capture valuable insights, a carefully constructed survey. This was designed to address the overall user experience, the ease of navigation, the clarity of the instructions, the performance of the tool, and the relevance of the questions to their correspondent topics.

The survey comprised a diverse array of quantitative and qualitative questions, care- fully crafted to ascertain the user's experiences, thoughts, and viewpoints. Quantitative inquiries delved into topics like the ease of navigation, how well the introductory video explained the quantum threat, whether the questions were logically divided into different sections, and the usefulness of the final report generated. To complement the quantitative aspects, the qualitative questions explored areas like usability issues, question clarity, whether the recommendations generated were appropriate and relevant, potential areas for improvement, and any noteworthy omissions in the questions relevant to quantum threats.

After receiving the survey results, it was identified that the first quantitative question could induce the survey's participants in error. This question was to rate the ease of navigation in the web application on a scale from 1 to 5. Usually, in this type of question option 1 means poor navigation, while option 5 would mean very good navigation, just like how the following quantitative question is configured. Unfortunately, in this case, it had opposite meanings, which might have led some survey participants to answer option 5, while intending to respond that the ease of navigation was good.

Overall, the survey's results illuminated both successes and opportunities for refinement in the application. After both surveys, the collected data was comprehensively analysed to obtain points of improvement. This analysis consisted of aggregating the quantitative responses to identify trends and perform a thematic analysis of the qualitative feedback. This combined approach improved the overall understanding of the web application's strengths and the areas that could be further improved.

The first survey was conducted based on a prototype of the web application, and the initial version of the set of questions. To conduct the survey, each participant was provided with the link to the deployed web application and a link to the survey they were supposed to fill out. Along with these, participant information documents were provided to inform the participants about the project's work being done, along with the right to withdraw their consent and halt their participation in this survey.

Regarding the web application, the feedback received in this initial survey was very positive. The overall user interface and responsiveness of the web application were praised for their clarity and ease of navigation. Regarding possible improvements, the main areas mentioned were the final risk score displayed on the results page, which was deemed too granular, as it is hard to quantify exactly the level of risk in a precise number from 0 to 100, and also because this would vary depending on each company or organisation sector and their opinions on what holds more value or importance. Besides this point, it was also mentioned that the long list of recommendations that the user receives on the results page could lead to a feeling of being overwhelmed with the sheer quantity of information provided in the case that there are too many recommendations. Based on feedback received, the web app deployment also had to be changed from the hosting service provider that was being used at the time, which was Render, to the university-based service instead.

Other feedback included in the survey was related to the introductory video in the application that aims to explain what the Quantum Threat is, which was very positive, emphasizing the idea that it clearly conveys the intended message. The sections created to break up the questions into logical compartments were also praised, in conjunction with the usefulness of the recommendations.

Regarding areas for possible improvement, the main topics discussed were the simplification and clarification of some questions, especially the ones regarding data storage, which could be difficult to interpret, the relevance of some IoT questions was also questioned, and the lack of clarity in a few recommendations, specifically the ones regarding the web browsers version, which did not specify which browser had to be updated and to which version. It was also mentioned that, while all recommendations displayed were useful, there should be more emphasis on the steps needed to transition into post-quantum cryptography or quantum attack resilient schemes. Furthermore, providing a detailed reason for each recommendation would be beneficial to the user's understanding.

Towards the end, a second survey, using the same questions but an updated version of the web application, was conducted to gather more feedback on the previously improved version. Regarding the web application, the feedback was mostly positive. The main areas mentioned for improvement were some of the components in the results page that are further discussed below.

The positive feedback areas mentioned in this second survey were mainly identical as mentioned previously in the first survey, including but not limited to the ease of navigation, the competent user interface and user experience, and the quality of the explanatory introduction video among others.

Further discussing areas for possible improvement, some of the topics mentioned included questions that would make more sense if they were multiple choice, instead of single choice. Examples include the question regarding what operating systems the company uses, the relevance of some questions was also pointed out, like the Bluetooth question, there was concern that some recommendations might be too granular, so it would be beneficial to only show the most important ones. Besides this, it was also suggested to decompose the recommendations list into separate sections, in the same way questions were divided into sections, this would result in improved clarity and a better understanding of which areas inside the organisation require the most improvements. Additionally, the insertion of extra answers to certain questions was also suggested, like adding the Safari option to the question regarding what web browsers are being used, to cover the most widely used ones. Another concern included the rewording of a couple of questions to facilitate a better understanding of what is being asked. The addition of follow-up questions regarding email server security protocols would deeply enhance our understanding of the organisation's security protocols in place for email communication. Finally, it was also suggested to add links to relevant basic resources that could aid organisations with their quantum-safe migration efforts. This could be added, as a separate section in the results page, or accomplished by simply expanding the existing recommendations with this information. The feedback gathered, laid the groundwork for the iterative improvements applied.

### 4.2. Design Improvement

Based on the feedback received from the first survey, many improvements were applied to the application. Regarding the web application, there were 2 main changes, both were on the results page. The first change included the final risk score number, which was replaced with a risk category. This risk category can be either low, medium, or high. The final numerical risk score will still be used to define what risk category should be used, but it will no longer be displayed to the user.

Regarding the boundary values of the categories, in terms of risk value, and taking into account that the maximum number is 100, if the total risk is lower than 34, then the risk category displayed will be low. If the risk value is higher than 59, then the risk category will be high. Otherwise, the risk category will be medium, so this will be the case when the risk value is equal to or anywhere between 34 and 59. Furthermore, the other change implemented was reducing the number of recommendations shown to the user on the results page. To accomplish this, a limit of 5 recommendations was implemented, for the results page to only display 5 recommendations or less. Nonetheless, if the user wants to have access to the full detailed list of recommendations, a see more button was added to allow this functionality.

Concerning the questions, the main changes implemented were related to simplifying and clarifying some questions, that were previously hard to understand. This included questions related to storage solutions, and discarding the IoT questions, which were found to not be relevant. Along with those changes, the recommendations used for the web browsers were also updated to directly mention on the recommendation which browser should be updated. Lastly, to cover the feedback regarding more quantum- specific recommendations and the inclusion of reasons behind each recommendation, some recommendations were updated to include this requested information.

Based on feedback from the second survey, the main changes were moving the graph displayed on the results page, into the start page, since it complements well the explan- atory video displayed there, and also because if it is displayed on the results page, it could confuse the user into thinking it is based on the user's answers, which it is not. Another change was adding into the questions page, a field that indicated the question type, so whether the current question displayed is single choice or multiple choice. This was done to indicate to the user how many answers they could select on a given question.

Besides this, the other change was adding a small paragraph explanation in the results page regarding the user's risk category, to better explain to the user what this risk category means.

Concerning the improvements made to questions, in this second iteration, the first change was adding links to relevant resources that could aid the company in its transition to post-quantum, instead of creating a specific section for these resources in the results page. This was accomplished by adding a link to a relevant resource in certain recommendations, like the one where the use of Slack is recommended. Besides this, some questions like the one regarding what operating system is used by the company devices, were changed to be multiple choice type, since the company environment might include different operation systems. While the suggestion to break up recommendations into different sections was not followed, the ordering of the recommendations displayed was improved again, to display the most relevant or urgent recommendations towards the top. Another implemented change was to expand the answer choices in some questions, to cover a broader range of commonly used software or protocols. An example of this was adding Safari to the possible answers to the question regarding what browsers are used by the organisation. While the extra suggested questions regarding email server security were not included, the questions and recommendations that received feedback mention- ing any difficulties in understanding them, were reworded and updated to address this issue.

Furthermore, some other improvements include allowing the user to click the Quantum- Watch logo on the top left of the header to be redirected back to the start page, making the help button inside each question only appear if there is help data available for it, and creating an About page describing the project.

## 5. Conclusions

In conclusion, a significant divide exists within organizations regarding the understanding and preparedness for the potential of quantum computers to compromise current encryption algorithms. This project aimed to bridge this gap by developing an evaluation tool for cybersecurity professionals to assess organizational vulnerabilities to quantum computing. Through extensive research and the creation of a web application, the project sought to raise awareness of these emerging threats and provide recommendations for securing systems against them.

The research findings highlighted that common asymmetric encryption algorithms are vulnerable to quantum computers due to Shor's algorithm, which accelerates the deciphering process for symmetric encryption. In response, standards bodies like NIST are developing quantum-resistant cryptographic protocols. The prototype tool created during this project was publicly deployed and reviewed by cybersecurity experts, receiving largely positive feedback for its effectiveness in illustrating the severity of quantum threats and evaluating quantum risks. Despite suggestions for simplifying questions and expanding coverage in certain areas, the tool was highly praised, particularly by Dr. Michael Murphy from Arqit, who commended it for exceeding expectations in design and quality. The survey results underscored the tool's capacity to educate users about the severity of quantum computing threats, suggesting that even simple educational materials can significantly address ignorance and apathy in cybersecurity. While feedback indicated that some technical questions were complex, the prototype successfully introduced unaware users to quantum threats and guided them towards available quantum-safe solutions. This project serves as a bridge, connecting users with solution providers and contributing to a sustainable future with robust post-quantum cryptographic security infrastructures. The prototype also has the potential to influence future research on post-quantum migration by helping non-expert users understand the associated risks. However, the project faced several challenges, including the absence of similar tools in the market, which complicated the quantitative assessment of the prototype. The evolving project requirements necessitated a flexible approach, occasionally at the cost of previous progress, but ultimately aligned product development with client needs, resulting in high satisfaction. The rapidly evolving field of quantum computing also posed challenges in verifying research data accuracy, especially given the team's initial lack of expertise. Weekly deliverables required precise task allocation and accountability within tight deadlines, which were managed through effective project management strategies. Looking ahead, future work could enhance user experience by improving accessibility and refining recommendation accuracy. Adding an information page summarizing the latest trends in quantum computing cryptography, integrating share buttons, and visualizing escalating quantum threats with interactive graphs could promote greater awareness within organizations. Despite modern browsers' default translation capabilities, a dedicated multi-language support feature might be necessary for global accessibility. Technically, implementing machine learning models could prioritize recommendations more effectively. For example, reinforcement learning could solicit user feedback to improve recommendation relevance and accuracy over time. Generative artificial intelligence could further enhance the tool's accessibility, especially by enriching the help section of the questionnaire to improve response accuracy and overall user experience. This remains a promising area for future research. Finally, incorporating more variables in the risk calculation could yield a more accurate assessment. Referencing the Common Weakness Enumeration database could provide a more precise approach to assessing breach likelihoods. However, this added complexity must be navigated carefully. This project aspires to enable a quantum-safe future before vulnerabilities become critical, aiming to enhance the overall security posture among organizations worldwide.

# Appendices

## Appendix 1: Questions

The complete list of all the questions used in the assessment tool is listed in Table A.1 alongside the recommendations associated with those questions if answers are selected that are not ideal for post-quantum security. The question numbers and their associated answers can be found in the Table A.2

Table A.1: Complete List of Assessment Questions and Recommendation

| Number | Question | Recommendation |
|---|---|---|
| 1.1 | Is your company in one of the following sectors? | |
| 1.2 | Does your company handle any of the following data types? | |
| 1.3 | Do you collect any personal information of EU citizens? | |
| 1.4 | How often does your organization conduct security audits? | Security Audits should be carried out at least twice a year to make sure protections are up to date and vulnerabilities aren't left exposed for extended periods of time. |
| 1.5 | How often does your organization carry out data backups? | Backups of sensitive data should be performed daily, and offsite backups should be created at least weekly to prevent data loss. |
| 1.6 | Have you completed an inventory of your cryptography in the last two years? | A cryptographic inventory should be taken of the whole business to establish which algorithms are in use and where. This allows an organization to be crypto-agile - able to act quickly to replace algorithms that may be broken, as well as making sure that when a post-quantum transition occurs there is a central location to track the progress and maintenance of the transition. |
| 1.7 | How often do your employees receive cyber security training? | Employees should be given cyber security training once every six months and all new employees should be trained before working with company or customer data. |

| 1.8 | Does your organization have a cyber security team? | A cyber security team should be created to monitor, update and manage the networks and devices of your business - they can potentially prevent at- tacks entirely or mitigate damages to the business significantly. |
| --- | --- | --- |
| 1.9 | Does your organization have a Cryptographic Center of Excellence frame- work in place? | A Cryptographic Centre of Excellence framework could be implemented to spread responsibility and awareness of the importance of encryption across all levels of the company. |

| 1.10 | Does your organization allow employees to work from home? | Any company data transferred off site should be utilizing properly encrypted channels such as Virtual Private Networks preferably which use post-quantum encryption algorithms. |
|---|---|---|
| 1.11 | What type of data flows through the network channel when working remotely? | |
| 1.12 | If you have proprietary cryptographic protocols in place within the organisation are they publicly available or reviewed regularly? | Make sure any proprietary or private cryptography is regularly reviewed to make sure it meets modern stand- ards - it should also be reviewed against quantum algorithms such as Shor's and Grover's. Cryptography algorithms are best verified by releas- ing to the public for widespread re- view and testing. |
| 1.13 | Are there plans to migrate to post-quantum encryption algorithms? If so, what is the timeline? | We recommend creating a network & security plan to ensure a timely and secure transition to quantum resist- ant algorithms. We recommend exploring the Arqit QuantumCloud tool to maintain a quantum key infrastructure. |
| | **VPNs and Communication** | |
| 2.1 | What level of security does your WIFI use? What is the type of your WiFi? | Data in transit between a device and a router is highly vulnerable to capture and cracking by quantum threats. Make sure the network is up to date with the latest security versions Upgrade Wi-Fi type. / Use fibre-optic cables and monitor for taps (allows detection without actor looking at the data content) - using Optical Time Domain Reflectometer (OTDR) for example. |

| 2.2 | How often are your organisation's devices reviewed for compatibility with evolving encryption standards? | Devices company-wide should have their specifications known and reviewed whenever encryption standards are updated to check if the hard- ware is capable of handling modern demanding encryption algorithms and if they are not they should be replaced and upgraded. |
|---|---|---|
| 2.3 | Do you utilize a VPN anywhere in your company network? | |
| 2.4 | Do you utilize a VPN service for employees connecting to company data from outside a place of work? | A VPN should be used for any employees working from home that handle company data - the VPN should utilize quantum resistant algorithms (AES-256, NIST approved asymmetric algorithms). |
| 2.5 | Do you utilize a VPN service when communicating between company offices? | A post-quantum VPN should be used even between company properties - the communication link over the inter- net will expose data in transit which could be collected and stored for later decryption by a quantum threat. |
| 2.6 | Which VPN provider are you using? | We would recommend the use of a post-quantum VPN to encrypt and protect company data whenever it is communicated between premises and out of office employees - i.e Arqit Fortinet Network Adaptor |
| 2.7 | Which protocol does your VPN use? | We suggest switching vpn provider to one utilising any of the following: OpenVPN, IKEv2/IPsec, SoftEther, SSTP, L2TP/IPsec, ChaCha20, AR-QIT Fortinet Network Adaptor |
| 2.8 | Does your infrastructure use any of these communication tools? | We would recommend the use of Slack for internal communication over similar services such as email due to it's publc enhanced security suite - 256 AES-GCM. We would also recommend the use of Signal due to its post quantum key exchange algorithm. https://arc.net/l/quote/xrhwcmit |
| 2.9 | Does your company operate its own email server? | |
| 2.10 | Which encryption protocols do you use for storing emails on your server? | We recommend implementing post-quantum end to end encryption in all company emails where possible. |
| 2.11 | How does your organization monitor and secure the email transmission protocols, specifically SMTP/S? | |
| 2.12 | Does your organization implement additional security measures on top of those provided by your email service provider? | |
| 2.13 | How are VoIP communications secured within your network to prevent eavesdropping or data interception? | If VoIP encryption is enabled check the algorithm used by your VoIP system is quantum resistant and NIST approved. |
| | **Locally Operated Networks** | |
| 3.1 | Does your organisation have security cameras (CCTV, IPcamera network), and if so are they connected to the internet? | Most on-premises CCTV or sensor networks do not need to be accessible over the internet, and should be locally routed to those monitoring them. This will prevent any access or data capture by a remote client and reduce the threat surface area. |
| 3.2 | Is there an Intrusion Detection/Prevention Systems in place? | An IDS/IPS is still useful in a post-quantum world - if encryption if broken it will still be able to detect intrusions made through use of statistics such as network flow metrics and access log trends. |

| 3.3 | Does your organisation utilise net- work segmentation? | Network segmentation should be implemented rigorously across any company network - sensitive data servers or storage devices should be separ- ate from each other within reason and from the rest of the company network with proper security control between them. Segmentation can reduce at- tack impact by preventing the spread of a security incident beyond one seg- ment of the network. |
|---|---|---|
| 3.4 | If your business uses custom/intern- ally developed tools indicate which encryption method(s) they use: | Any custom or bespoke tools and technologies should be utilizing known strong encryption technologies - make sure any such tools are considered for upgrading to the latest NIST post-quantum algorithms. |
| 3.5 | Do you have any network rules that prevent access to http websites? | Access to unencrypted websites (http) is not recommended due to the lack of encryption. Any data shared in the website is at high risk of being captured by third parties. We recommend implementing a web filter with your organisation's firewall to block- ing all outbound traffic on port 80 which will prevent standard HTTP connections. HTTPS should always be utilised over HTTP. |
| 3.6 | How do you manage files and data on the network? | Upgrade any FTP systems to SFTP - this utilizes the SSH protocol and large key AES encryption which is Quantum-resistant. FTPS utilizes TLS which is quantum-resistant but only when used with AES and ChaCha20 of significant key size. |
| 3.7 | How frequently are your organisation's network security settings (i.e TLS/SSL certificates, firewalls, access control) reviewed, patched, and updated? | Cyber security threats are constantly evolving and security settings should be constantly reviewed where pos- sible, ideally at least once a month to keep up to date with the current threats and vulnerabilities. |
| 3.8 | What local storage do you use for holding sensitive data? | Using an on-premises server or cloud deployment for accessing company data is recommended so company data isn't transported over the inter- net and potentially captured for later decryption |
| 3.9 | What is stored in the solutions? | |
| 3.10 | Is this data accessible over the inter- net or through a VPN? | A Quantum-resistant VPN service should be implemented so that com- pany data is secure during transport and sensitive information cannot be captured now and potentially cracked later in a post-quantum world. |
| 3.11 | Which protocol do you utilize for re- mote desktop access? | We recommend using VNC for your remote desktop access with the secur- ity setting set to "AlwaysMaximum". You can change this setting by follow- ing the steps: "Open the VNC Viewer app, and navigate to File >Prefer- ences >Expert. Search for the En- cryption parameter and set the value to AlwaysMaximum." (taken from of- ficial web). Alternatively SSH is quantum-resistant with large AES key size. |
| 3.12 | If any of your devices use Bluetooth select the version they use. | Bluetooth should be avoided for sens- itive information transmission as the current cryptography implementation is not quantum resistant. AES-128 is not considered secure against a quantum threat and other communic- ation platforms should be used where possible that support post-quantum cryptography. |

| 3.13 | Which channels are applicable to your computer network? | We recommend connections are only made where appropriate - highly sensitive or confidential departments should not be linked to non-sensitive networks or those that are exposed to the public for example. |
|---|---|---|
| | **Software Information** | |
| 4.1 | What web browsers does your infrastructure include? | We recommend the use of Chrome as your default browser, due to its quantum resistance against quantum- related threats and the adoption of modern quantum safe implementations. |
| 4.2 | Is X browser version higher than Y (yes, no, same, don't know) | We recommend that you update your browser to the latest version. This will ensure you have the highest level of security available to your current browser such as the usage of TLS 1.3 and use post-quantum encryption add-ons if they are available. |
| 4.3 | What kind of data do those browsers access? | Out of date browsers could have vulnerabilities and could result in a data leak - they should be kept up to date with all the latest security update. |
| 4.4 | What OS do your devices use? | |
| 4.5 | Is X OS version higher than Y (yes, no, same, don't know) | Older versions of operating systems that have finished their support/se- curity cycle often have major vulner- abilities and will not be updated to protect against future threats such as in a post quantum scenario. Some operating systems might not be able to support key technologies such as TLS and certain cryptography librar- ies. All OS's should be kept up to date and security patches deployed frequently. Before support is dropped for an OS all services should be trans- ferred elsewhere or the OS upgraded. |
| | **Cloud** | |

| 5.1 | What cloud infrastructure does your company have? | |
|---|---|---|
| 5.2 | *If public or hybrid* What public cloud services does your company infrastructure include? | |
| 5.3 | *If private or hybrid* What encryption does your cloud private service have? | Client Side Encryption with a Quantum resistant algorithm offers the highest level of security for data stored in the cloud as the data is encrypted before transit by the client machine. |
| 5.4 | What usage does your cloud infrastructure cover? | We recommend particularly sensitive data is not stored on a cloud solution unless strong preventative measures are in place to control access and post-quantum encryption algorithms are implemented to prevent data capture when interacting with the cloud service. |
| 5.5 | What protocols are used when connecting from internal to external cloud environments? | Lattice-based cryptography protocols are generally recommended when dealing with quantum computing threats. CRYSTALS-Kyber is a key encapsulation mechanism (KEM) we recommend using when connecting to different cloud environments |
| 5.6 | **Security** | |
| 5.7 | Does your company utilize digital signatures, e.g in contracts, transactions or email? | We recommend switching to one of the quantum resistant digital signatures recommended by NIST, like: CRYSTALS-Dilithium, FALCON, and SPHINCS+. Conventional digital signature technologies will not provided assurance of authenticity in a post-quantum world as they often rely on vulnerable asymmetric algorithms such as RSA and elliptic curve cryptography. |

| 5.8 | Does your company utilize RSA across any of its internal or external applications? | Asymmetric encryption is highly susceptible to quantum threats, and RSA is broken by quantum threats due to Shor's algorithm and the capabilities of a quantum computer. RSA should be upgraded to a NSIT approved asymmetric post-quantum cryptography algorithm such as Crystals-kyber. |
|---|---|---|
| 5.9 | What is the minimum TLS version allowed for HTTPS traffic? | Using an out of date TLS version could put you at more risk to quantum threats as their security suite may be weak or ineffective against quantum computers. TLS 1.3 is the latest version and we would recommend supporting and requiring this version everywhere. |
| 5.10 | Are any Post-Quantum enhancements utilized in your TLS communications? | Integration of any post-quantum enhancements as they get released can help increase quantum threats that can target TLS and data-in-transit. Tests have already been run such as KEMTLS (Cloudflare 2021) which show post-quantum TLS security in action and its capability. |
| 5.11 | What types of security controls are in place to manage traffic on your network? | Security measures such as access monitoring, access control lists and firewalls are integral to maintaining the integrity of a network and are still very relevant in a post quantum world. We recommend making sure each of these measures is in place throughout your network at a minimum: Access control lists, firewalls, Intrusion Prevention Systems, regular penetration testing. |
| 5.12 | Does your cloud infrastructure implement centralized logging and resource monitoring? | Logging and resource monitoring of a cloud network can help with prevention and mitigation prospects against any threats. |

| 5.13 | How does your organization currently create and manage crypto- graphic keys? | We recommend the usage of key management services like the cloud-based solution Arqit QuantumCloud to securely manage organization keys. Alternatively hardware devices are available but need configuration and maintenance. |
|---|---|---|

Table a.2: The Assessment Tool answer table

| Number | **Answer 1** | **Answer 2** | Answer 3 | Answer 4 | Answer 5 |
|---|---|---|---|---|---|
| | **General Information** | | | | |
| 1.1 | Class 1: Government, Critical Infrastructure, Military & Defense | Class 2: Financial Sector, Healthcare | Other | | |
| 1.2 | Class 1: Classified Data | Class 2: Personal Information/Financial Data/Health Information | None of the Above | | |
| 1.3 | Yes | No | | | |
| 1.4 | Biannually | Yearly | Less than once a year | | |
| 1.5 | Daily | Weekly | Monthly | Less than once a Month | |
| 1.6 | Yes | No | | | |
| 1.7 | Every 6 months | Yearly | No Regular Training | | |
| 1.8 | Internal or External Cyber Security Team | No Cyber Security Team | | | |
| 1.9 | Yes | No | | | |
| 1.10 | Yes - through an Internal VPN Connection | Yes - Directly over Internet | No | | |
| 1.11 | Customer/Client Information | Business/Employee Communications | Financial Data | None of the Above | |
| 1.12 | Yes - Publicly Available & Tested | Yes - Privately Reviewed | Yes - Not reviewed | No | |
| 1.13 | Yes - ASAP | Yes - In the next 5-15 years | No | | |

|     | **VPNs and Communication** |     |     |     |     |
| --- | --- | --- | --- | --- | --- |
| 2.1 | WPA-3 Enterprise or WPA with AES 256 | WPA2, WPA, WEP, WPA3 Personal or other | Other |     |     |
| 2.2 | Every year or less | Every couple years | Never |     |     |
| 2.3 | Yes | No  |     |     |     |
| 2.4 | Yes | No  |     |     |     |
| 2.5 | Yes | No  |     |     |     |
| 2.6 | Post-Quantum VPN (i.e Arqit Fortinet Network Adaptor) | Cisco AnyConnect, Palo Alto Networks Global Protect, Fortinet FortiClient, Citrix Gateway, OpenVPN, Perimeter 81, IBM Security MaaS360 VPN, Google Cloud VPN, Azure VPN Gateway | Other |     |     |
| 2.7 | OpenVPN, IKEv2/IPsec, SoftEther, SSTP, L2TP/IPsec, ChaCha20 (all use AES-256, except ChaCha20) | Other | Other |     |     |
| 2.8 | Microsoft Teams | WhatsApp (claims E2EE, no algorithm officially announced) - can be grouped with other imo | Email | Other |     |
| 2.9 | Yes | No  |     |     |     |
| 2.10 | TLS/SSL | End to end advanced quantum secure encryption | None | | |
| 2.11 | SMTP | Sender Policy Framework (SPF) | DomainKeys Identified Mail (DKIM) | | |
| 2.12 | Additional encryption and access controls. | Employee training | No | | |
| 2.13 | VoIP service isolation | VoIP traffic encryption | VoIP device authentication | | |
| | **Locally Operated Networks** | | | | |
| 3.1 | Yes | No | | | |
| 3.2 | Yes | No | | | |
| 3.3 | Yes | No | | | |
| 3.4 | NIST-Approved Post Quantum Algorithms | RSA, AES-128 or below, | Other | | |
| 3.5 | Yes | No | | | |
| 3.6 | SFTP | FTPS | FTP | Other | |
| 3.7 | Every week | Monthly | Yearly | Never | |
| 3.8 | Local device storage | On-site Cloud storage | No On-Premises Storage | | |
| 3.9 | Sensitive Information | Customer Personal Information | Public Information | None of the above | |
| 3.10 | Internet | Through VPN | Only Locally | | |
| 3.11 | VNC | SSH (uses symmetric encryption and you can choose) | PCoIP (SALSA20/12-256, AES-GCM-128, and AES-GCM-256. | RSA certificate based, RDP, Telnet, or Other (not quantum safe) | |
| 3.12 | Bluetooth LE | Bluetooth 3 or lower | | | |

| 3.13 | Office to Public Cloud | Data Centre to Data Centre | IoT to Data Centre | Employee Device to Internal Server Share | |
|---|---|---|---|---|---|
| | **Software In-formation** | | | | |
| 4.1 | Google Chrome | Firefox | Microsoft Edge | Safari (or Other - answer 5) | |
| 4.2 | Dependent on answer: Chrome (Version 70-122 TLS 1.3 support), Firefox (Version 63-123 TLS 1.3 support), MS Edge (Version 79-119 TLS 1.3 support), Sa- fari (version 14 and above has TLS1.3) | No | | | |
| 4.3 | Company | Client | Public | Other | |
| 4.4 | Windows | MacOsS | Linux | | |
| 4.5 | macOS 10.15 and later TLS 1.3 support (+AES, Windows 10 (File encrypted with AES-256 bit and later TLS1.3 support, Linux version 8 and later TLS 1.3 support, also uses AES-256 bit | macOS 10.15 and earlier no TLS 1.3 support, Windows 10 and earlier no TLS1.3 support, Linux version 8 and earlier no TLS 1.3 support | | | |
| | **Cloud** | | | | |
| 5.1 | Public | Private | Hybrid | | |
| 5.2 | Google Cloud | Amazon Web Services (AWS) | Azure | Other | |
| 5.3 | Server Side Encryption (SSE) | Client Side Encryption (CSE) | Other | | |
| 5.4 | Data and File Storage | Data Processing | Hosting | Other | |
| 5.5 | TLS/SSL | IPSEC | CRYSTALS-Kyber | Other | |
| 5.6 | **Security** | | | | |
| 5.7 | Yes | No | | | |
| 5.8 | Yes | No | | | |
| 5.9 | 1.1 | 1.2 | 1.3 | | |
| 5.10 | Yes | No | | | |
| 5.11 | Access control lists (ACLs) | Firewalls | Intrusion prevention systems (IPS) | Penetration testing | |
| 5.12 | Yes | No | | | |
| 5.13 | Hardware Security Modules (HSMs) | Custom In-House Solutions | Cloud-Based Key Management Services | Open-Source Key Management Solutions | Blockchain-Based Key Management |